\documentclass[prb,final,twocolumn,superscriptaddress,showpacs]{revtex4}
\usepackage{graphicx}

\begin{document}

\title{Magnetoresistance scaling in the layered cobaltate Ca$_3$Co$_4$O$_9$}

\author{P. Limelette}
\affiliation{Laboratoire LEMA, UMR 6157 CNRS-CEA, Universit\'e F. Rabelais, UFR Sciences, Parc de Grandmont, 37200 Tours, France}
\author{J. C. Soret}
\affiliation{Laboratoire LEMA, UMR 6157 CNRS-CEA, Universit\'e F. Rabelais, UFR Sciences, Parc de Grandmont, 37200 Tours, France}
\author{H. Muguerra}
\affiliation{Laboratoire CRISMAT, UMR 6508 CNRS-ENSICAEN et Universit\'e de Caen, 6, Boulevard du Mar\'echal Juin, 14050 CAEN Cedex, France}
\author{D. Grebille}
\affiliation{Laboratoire CRISMAT, UMR 6508 CNRS-ENSICAEN et Universit\'e de Caen, 6, Boulevard du Mar\'echal Juin, 14050 CAEN Cedex, France}

\date{\today}

\begin{abstract}
\vspace{0.3cm}
We investigate the low temperature magnetic field dependences of both the resistivity and the magnetization in the misfit cobaltate Ca$_3$Co$_4$O$_9$ from 60 K down to 2 K. 
The measured negative magnetoresistance reveals a scaling behavior with the magnetization which demonstrates a spin dependent diffusion mechanism. 
This scaling is also found to be consistent with a shadowed metalliclike conduction over the whole temperature range. 
By explaining the observed transport crossover, this result shed a new light on the nature of the elementary excitations relevant to the transport.
\end{abstract}


\maketitle

\section{introduction}

Since correlated oxides as the superconducting cuprates or the manganites with their colossal magnetoresistance \cite{Imada98,Salamon01} have exhibited outstanding properties, the physics of strongly correlated electron materials has motivated an extensive research including both experimental and theoretical investigations \cite{Imada98,Georges96}. 
The hunt for new materials has led to the discovery of the layered cobaltates,  with striking properties \cite{Terasaki97}, a rich phase diagram \cite{Foo04} including superconductivity \cite{Takada03} and both metalliclike and insulatinglike characteristics. 
These compounds being composed of triangular sheets of cobalt atoms, they share in common with the high-temperature superconductor cuprates 
the fascinating physics combining the strong correlations and the geometrical frustration in two dimension \cite{haerter06}.

Also, most of the cobaltates display at room temperature a metallic resistivity and a large thermopower allowing for thermoelectric applications.
While it is known that either the entropy of localized spins \cite{Koshibae00,Maignan03} or the electronic correlations \cite{Palsson98} can increase thermopower, it has been suggested despite their antagonism that both of them seem to contribute in cobaltates \cite{LimelettePRL06}. 
The latter interpretation being still highly debated, the interplay between magnetism and transport properties appears as a very challenging issue to understand the underlying physics in the cobaltates\cite{xiang07}.

In addition, various misfit cobaltates exhibit complex transport properties as metal-insulator crossover and large negative magnetoresistance 
\cite{Masset00,Maignan03,yamamoto02} which raise the question of the nature of the elementary excitations.
In this context, we report in this article on an experimental study of the single crystal magnetoresistance and magnetization in the misfit cobaltate 
Ca$_3$Co$_4$O$_9$ in order to connect both of them and to discuss its metallicity.

The studied single crystals were grown using a standard flux method \cite{Xia2005}, with typical sizes of the order of  4$\times$2$\times$0.02 mm$^3$.
Similarly to Na$_{x}$CoO$_2$ \cite{Terasaki97}, the structure of the incommensurate cobaltate [CoCa$_{2}$O$_{3}$]$^{RS}_{0.62}$CoO$_{2}$ 
(abbreviated thereafter CaCoO) contains single [CoO$_2$] layer of CdI$_2$ type stacked with rocksalt-type layers labeled $RS$ instead of a sodium deficient layer.
One of the in-plane sublattice parameters being different from one layer to the other \cite{Masset00}, the cobaltate CaCoO has a misfit structure as in most related compounds \cite{Maignan03}.

\section{measurements}

\begin{figure}[htbp]
\centerline{\includegraphics[width=0.95\hsize]{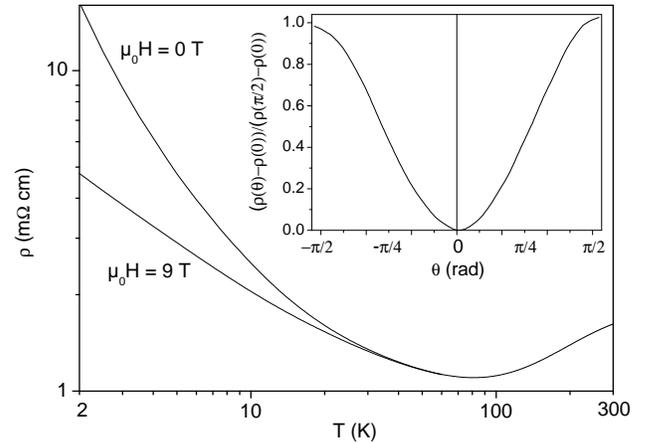}}
\caption{Temperature dependences of the single crystal in-plane resistivity $\rho$ at 0 T and 9 T, the magnetic field being parallel to the CoO$_{2}$ planes. The inset displays the variation of the normalized magnetoresistance with the angle between the CoO$_{2}$ planes and the 9 T magnetic field at T=60 K.
}
\label{fig1}
\end{figure}

As frequently observed in misfit cobaltates \cite{Masset00,Maignan03,yamamoto02}, the temperature dependence of resistivity displayed in Fig.~\ref{fig1} exhibits a transport crossover from a high temperature metalliclike regime to a low temperature insulatinglike behavior below nearly 70 K. 

Beyond this crossover, two interesting features are worth noting. 
First, the values of the resistivity remain of the order of the Mott limit \cite{Mott79} over the whole temperature range, \textit{i.e.} nearly 3 m$\Omega$ cm in this compound. 
This means that the electronic mean free path is of the order of the magnitude of the interatomic spacing as in a bad metal in the vicinity of a Mott metal-insulator transition \cite{Georges96,Limelette03PRL} for example. 
Also, by showing a large negative magnetoresistance, the low temperature insulatinglike behavior in Fig.~\ref{fig1} tends to be suppressed by the application of a  magnetic field. 
As a consequence of the two previous characteristics, the latter regime does not seem to result from a gap opening in the one-particle electronic spectrum 
as it has early been proposed \cite{sugiyama02}.
As a matter of fact, such a scenario would require a gap much more smaller than the crossover temperature, which invalidates an activated conduction and leads to unrealistic Coulomb interaction energy \cite{Singh}. 
Then, in order to investigate this particular low temperature regime, we have performed both magnetoresistance and magnetization measurements in CaCoO.

\subsection{Magnetoresistance}

First, we have measured  the single crystal in-plane resistivity $\rho$(H,T) using a standard four terminal method at constant temperature T as a function of an in-plane magnetic field $\mu_0$H, for which the negative magnetoresistance displayed in the inset of Fig.~\ref{fig1} is the largest. 

\begin{figure}[htbp]
\centerline{\includegraphics[width=0.95\hsize]{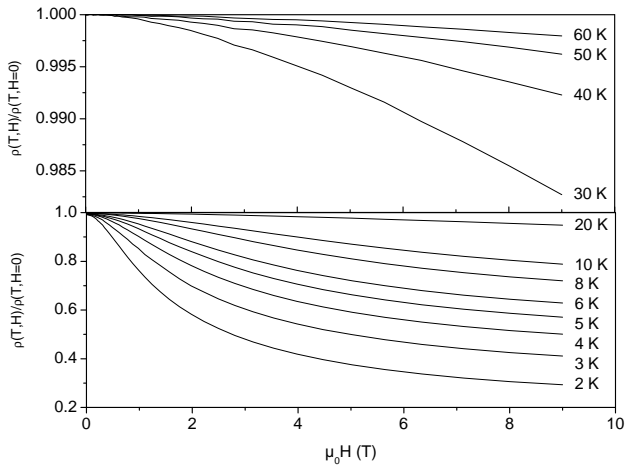}}
\caption{The upper and lower panels display the magnetic field dependences of the normalized resistivity $\rho$(T,H) over the whole investigated temperature range, the magnetic field being parallel to the CoO$_{2}$ planes.
}
\label{fig2}
\end{figure}

The data reported in Fig.~\ref{fig2} spans the T-range from 60 K down to 2 K with a large negative magnetoresistance reaching at the lowest temperature 70 $\% $ at 9 T.

\subsection{Magnetization}

\begin{figure}[htbp]
\centerline{\includegraphics[width=0.95\hsize]{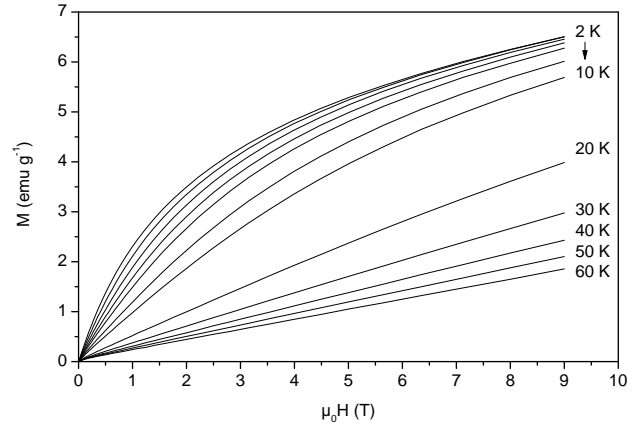}}
\caption{Magnetic field dependence of the single crystal in-plane magnetization up to 9 T over the whole temperature range, the magnetic field 
being parallel to the CoO$_{2}$ planes.}
\label{fig3}
\end{figure}

Also, the single crystal in-plane first magnetization has been measured using a magnetometer of a Quantum Design physical property measurement system on a sample of 3.2 mg. 
In order to be abble to compare these results with the magnetoresistance, the data set reported in Fig.~\ref{fig3} has been measured by applying an in-plane magnetic field.

As it is suggested in Fig.~\ref{fig3} by the sizeable increase of the first magnetization below nearly 20 K 
a magnetic ordering, already interpreted as a ferrimagnetic transition \cite{sugiyama02}, occurs within this temperature range. 
Although it is not displayed in Fig.~\ref{fig3}, we emphasize that hysteretic behavior has been observed below 20 K in agreement with the previously reported results \cite{sugiyama02}.

\section{scaling analysis}

\subsection{Low temperature regime}

\begin{figure}[htbp]
\centerline{\includegraphics[width=0.95\hsize]{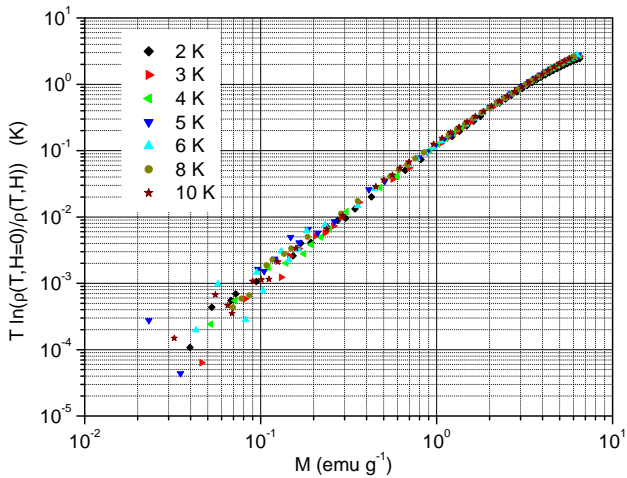}}
\caption{(color online). Scaling plot of the single crystal in-plane magnetoresistance as a function of the magnetization at low temperatures. 
}
\label{fig4}
\end{figure}

So, let's first focus on the low temperatures data, namely below 20 K. 
In this temperature range, Fig.~\ref{fig4} demonstrates that the magnetoresistance fully scales with the magnetization according an activated form which can be written following Eq.~\ref{eq0}.

\begin{equation}
\rho(T,H)=\rho(T,H=0) \mbox{\ \ } exp \left( - \alpha \frac{M^{\mu}}{T}\right)
\label{eq0}
\end{equation}

where $\alpha$ is a constant and the exponent $\mu$ equals approximately 2 as expected in the spin dependent transport mechanism proposed by Wagner \textit{et al}. \cite{Wagner98,Viret97} to explain the colossal magnetoresistance measured in the manganites.

\begin{figure}[htbp]
\centerline{\includegraphics[width=0.85\hsize]{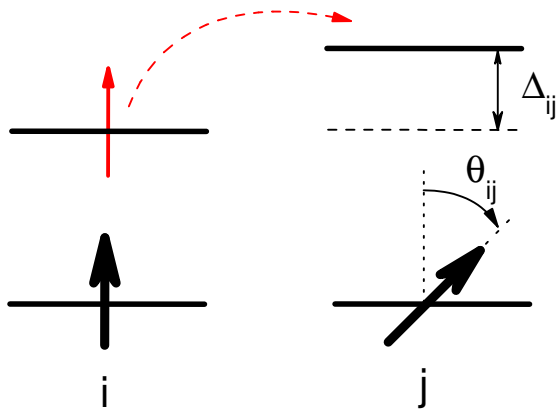}}
\caption{(color online). Schematic picture of the electronic energy levels involved in an elementary spin dependent hopping from the site i to the site j. 
The magnetic potential barrier $\Delta_{ij}$ results from the misorientation $\theta_{ij}$ between the quasiparticle spin s$_i$ (thin red arrow) and the localized spin $\sigma_j$ (bold black arrow) as explained in the text.
}
\label{fig5}
\end{figure}

This spin dependent transport mechanism is in fact an extension of a variable range hopping \cite{Mott79,efros75} to the case of magnetic disorder \cite{Wagner98,Viret97}. 
The main ideas which underlie this conducting process consist in a variable range spin dependent hopping due to the Hund's coupling -J$_H$\textbf{s}.\textbf{$\sigma$} between the quasiparticles spin \textbf{s} and the localized spin \textbf{$\sigma$}, with the Hund's coupling constant J$_H$. 
To basically illustrate this mechanism, let us consider the propagation of a quasiparticle with a spin s$_i$ from the site i to the site j as sketched in 
Fig.~\ref{fig5} with the involved spin polarized electronic energy levels.

Because s$_i$ points along the localized spin $\sigma_i$ direction, the misorientation $\theta_{ij}$ between $\sigma_i$ and the neighbor localized spin $\sigma_j$ leads to a magnetic potentiel barrier $\Delta_{ij}$=$\Delta$ (1-$\cos{\theta_{ij}}$)/2 with $\Delta$=2J$_H$ s$\sigma$. 
These magnetic potential barriers being distributed over the sample, they favor a hopping conduction of either activated or VRH type depending on the barriers distribution. 
Indeed, one might expect an activited conduction if magnetic correlations exist from one site to another (case of a weak magnetic disorder), or a VRH conduction if the paramagnetism persists down to the low temperatures (case of a strong magnetic disorder). 
Therefore, the reported activated behavior in Fig.~\ref{fig4} seems consistent with the presence of magnetic correlations in this temperature range in agreement with the magnetization data displayed in Fig.~\ref{fig3}.

Moreover, the previously mentioned misorientation can be expressed from the local magnetizations $\vec{M}_{i,j}$ at the two sites i and j  as $\vec{M}_i \cdot \vec{M}_j$=M$_S^2$ cos $\theta_{ij}$, with the saturation magnetization M$_S$. 
It follows that the local magnetic potentiel barrier is a function of local magnetizations. 
Then, an applied magnetic field which tends to align the localized spins, lowers the magnetic potentiel barrier and thus implies a negative magnetoresistance as observed in Fig.~\ref{fig1} and \ref{fig2}.

Let us for now discuss the magnetization exponent infered from the scaling plot in Fig.~\ref{fig4}. 
In the spin dependent transport mechanism proposed by Wagner \textit{et al.} \cite{Wagner98}, the local magnetizations M$_{i,j}$ are the sum of two different contributions: 
the Weiss magnetization M$_W$, which is common to both sites i and j, and the local corrections $\delta$M$_{i,j}$. 
The magnetic field dependent part of the average potentiel barrier $\langle \Delta_{ij} \rangle$ varying as $\langle \cos{\theta_{ij}} \rangle$, it  can then be related to the product $\langle \vec{M}_i \cdot \vec{M}_j \rangle$ and its following expansion: 

\begin{equation}
\langle \vec{M}_i \cdot \vec{M}_j \rangle = \left( M_W \right)^2 + 2 M_W \langle \delta M_i \rangle + \langle \delta M_i \: \delta M_j \rangle
\label{eq2}
\end{equation}

\noindent Three limiting cases can thus occur, involving one of the three terms in the previous relation.

 i) In a paramagnet where M$_W$=0, the first two terms in Eq.~\ref{eq2} vanish and $\langle \vec{M}_i \cdot \vec{M}_j \rangle$ is expected to scale as the third one, $\langle \delta M_i \: \delta M_j \rangle$=$\langle \delta M_i \rangle^2$, namely the square of the paramagnetic magnetization.\\
In the opposite limit, in a ferromagnetic state with M$_W \gg \langle \delta M_i \rangle$, $\langle \vec{M}_i \cdot \vec{M}_j \rangle$ may be  dominated by the first two terms.

 ii) Therefore, if the Weiss magnetization is constant, \textit{i.e.} both magnetic field and temperature independent (in a high magnetic field regime), $\langle \vec{M}_i \cdot \vec{M}_j \rangle$ should vary as the second term, namely $\langle \delta M_i \rangle$.

 iii) In a low magnetic field regime or in a soft ferromagnet, if M$_W$ is both magnetic field and temperature dependent but still higher than $\langle \delta M_i \rangle$, $\langle \vec{M}_i \cdot \vec{M}_j \rangle$ should then scale as the first term, namely M$_W^2$.

In particular, cases i) and iii) explain why the exponent $\mu$=2 is expected in the low magnetization regime, either correlated or not.

\subsection{High temperature regime}

Furthermore, in order to extend the scaling analysis to higher temperatures, namely above 10 K, let us consider that the resistivity results from two conduction mechanisms. 
The spin dependent one, $\rho_{sp}$, gives rise to the observed large negative magnetoresistance whereas the other, $\rho_{QP}$, is assumed to be weakly magnefic field dependent. Here, the initials QP stand for quasiparticles.
According to the Matthiessen rule, the experimental resistivity $\rho(T,H)$ can then be written following Eq.~\ref{eq1}. 

\begin{equation}
\rho(T,H) \approx \rho_{sp}(T,H) + \rho_{QP}(T)
\label{eq1}
\end{equation}

As a matter of fact, if we directly perform the same scaling as in Fig.~\ref{fig4} the data are systematically shifted downward suggesting an incorrect normalization at zero magnetic field and thus, an additional contribution to the resistivity weakly magnefic field dependent. 

\begin{figure}[htbp]
\centerline{\includegraphics[width=0.95\hsize]{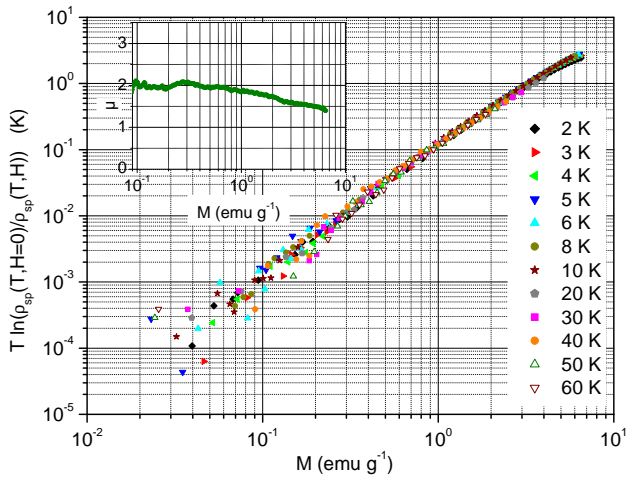}}
\caption{(color online). Scaling plot of the effective single crystal in-plane magnetoresistance 
$\rho_{sp}$(T,H) as a function of the magnetization over the whole temperature range. 
Here, $\rho_{sp}$(T,H) has been calculated by substrating to the experimental value a constant resistivity at each temperature above 10 K. 
To make sense, the substracted part is assumed to be weakly magnetic field dependent. 
The inset displays the variation of the exponent $\mu$ with the magnetization.}
\label{fig6}
\end{figure}

Therefore, the previous scaling can be extended in Fig.~\ref{fig6} over the whole temperature range by substracting the contribution $\rho_{QP}$ to the experimental resistivity up to 60 K, $\rho_{QP}$ being considered as a free parameter. 
Also, the inset of the Fig.~\ref{fig6} displays that the exponent $\mu$ equals 2 in the low magnetization regime, either correlated (case iii)) or not (case i)), 
and seems to slightly decrease above M$ \approx $ 1 emu g$^{-1}$ as suggested in the case ii).

\begin{figure}[htbp]
\centerline{\includegraphics[width=0.95\hsize]{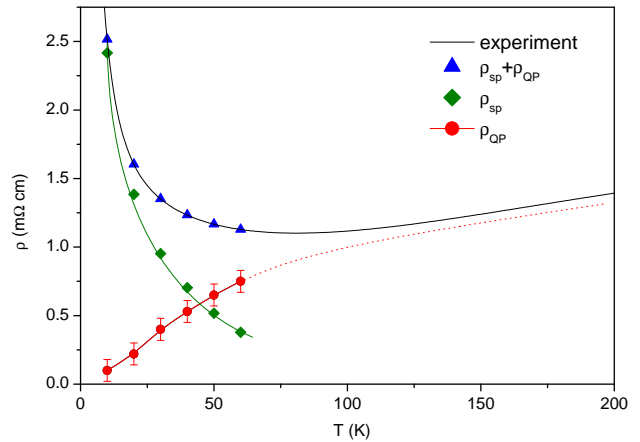}}
\caption{(color online). Temperature dependences of the experimental resistivity, the subsracted metalliclike part $\rho_{QP}$ and the deduced insulatinglike part $\rho_{sp}$. The dotted line is an extrapollation to recover the high temperature metalliclike behavior.}
\label{fig7}
\end{figure}

Interestingly, the T-dependence of the contribution $\rho_{QP}$  reveals in Fig.~\ref{fig7} a continuous metalliclike behavior down to the low temperatures. 
Since the metallic part of the magnetoresistance is usually very weak, the found component agrees with the initial assumption of an additional contribution weakly magnetic field dependent. 

Besides, both the T-dependence and the order of magnitude of  $\rho_{QP}$ are consistent with the metalliclike high temperature regime. 
As a result, the transport crossover around 70 K appears more as the temperature below which a diffusion mechanism predominates over the other rather than a consequence of a qualitative evolution of the system. \\

In other words, the electronic transport seems to originate from metallic or quasi-metallic conduction over the whole temperature range but shadowed at low temperatures by an efficient spin dependent diffusion mechanism. 
Further both experimental and theoretical investigations are now needed in order to understand the influences of the misfit structure as well as the frustration induced by the triangular lattice on this shadow metallicity and its connection with the thermopower.

\section{conclusion}

To conclude, we have investigated the low temperature magnetic field dependence of both the resistivity and the magnetization in the misfit cobaltate CaCoO from 60 K down to 2 K. 
The measured negative magnetoresistance reveals a scaling behavior with the magnetization which demonstrates a spin dependent diffusion mechanism in addition to a shadowed metalliclike conduction over the whole temperature range. 
This result provides a natural explanation of the observed transport crossover and shed a new light on the nature of the elementary excitations relevant to the transport.

\begin{acknowledgments}
It is a pleasure to acknowledge useful discussions with S. H\'ebert, R. Fr\'esard and Ch. Simon.
\end{acknowledgments}


\end{document}